\documentclass[conference]{IEEEtran}
\IEEEoverridecommandlockouts
\usepackage{cite}
\usepackage{amsmath,amssymb,amsfonts}
\usepackage{algorithm}      
\usepackage{algpseudocode}
\usepackage{graphicx}
\usepackage{textcomp}
\usepackage{xcolor}
\usepackage{caption}
\usepackage{subcaption}
\usepackage{balance}

\usepackage[pdftex]{hyperref}
\usepackage{tikz}

\newcommand\copyrighttext{%
    \footnotesize \textcopyright 2024 IEEE. Personal use of this material is permitted.
    Permission from IEEE must be obtained for all other uses, in any current or future
    media, including reprinting/republishing this material for advertising or promotional
    purposes, creating new collective works, for resale or redistribution to servers or
    lists, or reuse of any copyrighted component of this work in other works.
    DOI: \href{https://doi.org/10.1109/e-Science62913.2024.10678686}{https://doi.org/10.1109/e-Science62913.2024.10678686}}
\newcommand\copyrightnotice{%
    \begin{tikzpicture}[remember picture,overlay]
        \node[anchor=south,yshift=10pt] at (current page.south) {\fbox{\parbox{\dimexpr\textwidth-\fboxsep-\fboxrule\relax}{\copyrighttext}}};
    \end{tikzpicture}%
}

\def\BibTeX{{\rm B\kern-.05em{\sc i\kern-.025em b}\kern-.08em
    T\kern-.1667em\lower.7ex\hbox{E}\kern-.125emX}}
\begin{document}

\title{KS+: Predicting Workflow Task \\ Memory Usage Over Time
}

\author{
    \IEEEauthorblockN{Jonathan Bader\IEEEauthorrefmark{1}\IEEEauthorrefmark{3}, Ansgar Lößer\IEEEauthorrefmark{1}\IEEEauthorrefmark{2}, Lauritz Thamsen\IEEEauthorrefmark{4},
    Björn Scheuermann\IEEEauthorrefmark{2},
        and Odej Kao\IEEEauthorrefmark{3}}

    \IEEEauthorblockA{
        \IEEEauthorrefmark{3}
        \{firstname.lastname\}@tu-berlin.de, Technische Universität Berlin, Germany\\
    }
    \IEEEauthorblockA{
        \IEEEauthorrefmark{2}
        \{firstname.lastname\}@kom.tu-darmstadt.de, Technische Universität Darmstadt, Germany\\
    }
    \IEEEauthorblockA{
        \IEEEauthorrefmark{4}
        lauritz.thamsen@glasgow.ac.uk, University of Glasgow, United Kingdom\\
    }

}
\IEEEpubid{\makebox[\columnwidth]{*equal contribution \hfill} \hspace{\columnsep}\makebox[\columnwidth]{ }}

\maketitle
\copyrightnotice

\begin{abstract}
Scientific workflow management systems enable the reproducible execution of data analysis pipelines on cluster infrastructures managed by resource managers such as Kubernetes, Slurm, or HTCondor. 
These resource managers require resource estimates for each workflow task to be executed on one of the cluster nodes.
However, task resource consumption varies significantly between different tasks and for the same task with different inputs.
Furthermore, resource consumption also fluctuates during a task's execution.
As a result, manually configuring static memory allocations is error-prone, often leading users to overestimate memory usage to avoid costly failures from under-provisioning, which results in significant memory wastage.

We propose KS+, a method that predicts a task's memory consumption over time depending on its inputs.
For this, KS+ dynamically segments the task execution and predicts the memory required for each segment.
Our experimental evaluation shows an average reduction in memory wastage of 38\% compared to the best-performing state-of-the-art baseline for two real-world workflows from the popular nf-core repository.

\end{abstract}


\section{Introduction}\label{sec:INTRO}
As the amount of scientific data available increases, it is becoming impractical for scientists to analyze their datasets manually using sequences of custom scripts.
Scientists, therefore, increasingly rely on scientific workflow management systems (SWMS) to help them define, execute, orchestrate, and monitor their data analysis workflows~\cite{deelman2015pegasus,da2017characterization,baderWorkflowMonitoring}.
Such data analysis workflows typically consist of a set of interdependent analysis steps, called tasks, whose execution order is defined by data dependencies. 
Thousands of instances of the same task can be executed with different data inputs when executing a workflow on large datasets~\cite{da2013toward,callaghan2024using,tovar2022dynamic}.
It is impractical to run such a large number of tasks on a scientist's personal computer.
Thus, SWMS enable interoperability with various resource managers such as Slurm~\cite{yoo2003slurm}, HTCondor~\cite{thain2005distributed}, or Kubernetes~\cite{burns2016borg} to execute the data analysis workflow on a cluster infrastructure.

\begin{figure}[t!]
\center
\begin{subfigure}[t]{0.8\columnwidth}
   \raggedleft
  \includegraphics[width=1\columnwidth, trim={0mm 0mm 0mm 10mm}]{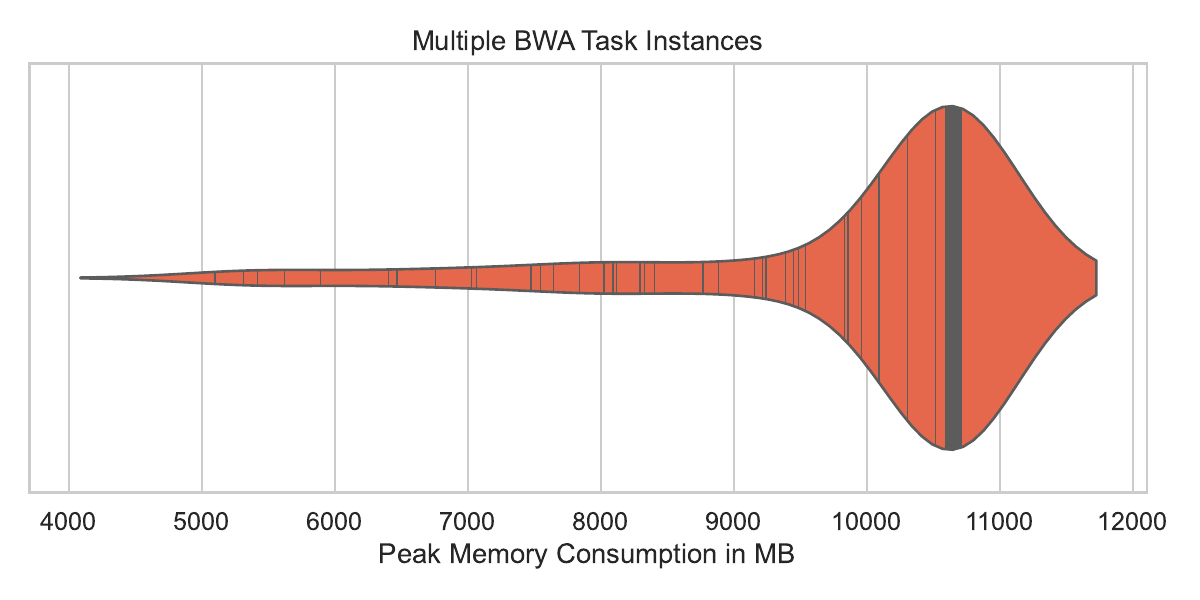}
  \caption{Distribution of peak memory consumption of multiple BWA task executions. Lines show observations.}
  \label{fig:distribution_peak_memory}
\end{subfigure}
\begin{subfigure}[t]{0.8\columnwidth}
  \raggedleft
  \includegraphics[width=1\columnwidth, trim={21mm 0mm 0mm 0mm}]{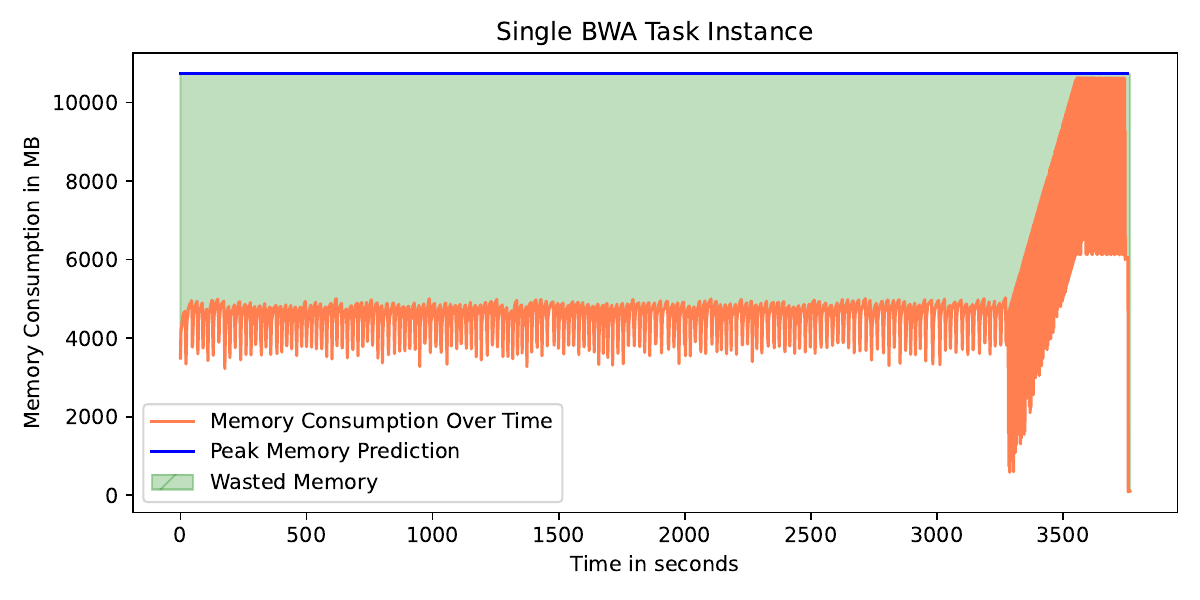}
  \caption{A BWA task's memory consumption over time.}
  \label{fig:memory_usage_over_time}
\end{subfigure}
\caption{Variability of memory consumption a) for the Burrows-Wheeler Aligner (BWA) task with different inputs and b) for a single BWA task over time during execution.}
\label{fig:eval_total}
\end{figure}%

Resource managers require memory usage limits for each task instance.
Exceeding these limits will result in costly bottlenecks or, more often, task termination~\cite{albahar2022schedtune, kintsakis2019reinforcement,chen2019augmented}.
Thus, scientists are effectively incentivized to overestimate resource consumption to avoid timely re-executions of tasks or even complete workflows.
However, requesting more memory than needed wastes precious cluster resources and limits the throughput on both a workflow and a cluster level.

The problem of setting accurate resource limits for tasks is further aggravated by the fact that the same task consumes different amounts of memory for different inputs, often depending on the size of the input data.
Figure~\ref{fig:distribution_peak_memory} shows the distribution of peak memory consumption for the Burrows-Wheeler Aligner (BWA) task — a popular tool for aligning short DNA sequences — when running the eager workflow~\cite{yates2021reproducible}.
It can be observed that the median memory consumption is around 10600~MB. 
However allocating the median memory usage for all BWA tasks would result in considerable memory wastage for half of the tasks, while the other half would fail. 
In addition, most tasks do not maintain a constant memory usage throughout their execution.
For example, a task may wrap multiple existing programs, resulting in varying memory consumption patterns as each of the programs is executed.
Even when a task wraps only one program, memory usage often fluctuates. 
A common example is a task that first loads input data into memory before processing it.
An instance of such non-static memory consumption can be seen in Figure~\ref{fig:memory_usage_over_time}.
Here, the BWA task uses about 5.1~GB of memory for about 80\% of its runtime, before the memory usage more than doubles to finally around 10.7~GB.
Simply allocating a fixed value of 10.7~GB would work, but would waste substantial memory as highlighted in green in the figure.

Currently, state-of-the-art workflow memory prediction methods only estimate a task's peak memory usage.
To do this, many methods use machine learning techniques, such as regression models\cite{witt2019feedback,witt2019learning}, reinforcement learning~\cite{bader2022leveraging}, or ensemble methods that combine multiple machine learning models~\cite{rodrigues2016helping,bader2024Sizey}.
Recently, we proposed the k-Segments method~\cite{baderDiedrichPredicting2023}, which predicts workflow task memory consumption over time by leveraging time series monitoring data.
Our method estimates a task's runtime, divides it into $k$ equally sized segments, and predicts the memory consumption for each segment based on the data input size.
However, as can be seen in Figure~\ref{fig:memory_usage_over_time}, equally sized segments are not necessarily able to model a task's memory consumption over time accurately.

In this paper, we present KS+, an extension of our previously proposed method, which works as follows:
First, KS+ dynamically sizes the segments and predicts peak memory values for each segment.
KS+ ensures that the prediction function is monotonically increasing, preventing task failures caused by reducing memory too early.
Subsequently, KS+ offsets the predictions to improve reliability.
In addition, if a task fails, KS+ adaptively adjusts the segments' position and the predicted memory before re-execution.

The contributions of this paper are:
\begin{itemize}
    \item We introduce KS+, which dynamically predicts the memory consumption of workflow tasks by dynamically estimating the position, length, and memory consumption for individual task segments. The source code is publicly available\footnote{github.com/dos-group/KSPlus}.
    \item We propose a novel task failure strategy for dynamic task memory prediction that re-adjusts not only the memory but also the segments for task re-execution.
    \item We evaluate our KS+ method on traces from two real-world workflows, using different amounts of training data, and show for example an average memory wastage reduction of 38\% compared to the original k-segments method and 51\% compared to the best-performing peak memory prediction baseline. 
\end{itemize}

\section{Approach}\label{sec:APPROACH}
\begin{figure}[t!]
\center
\includegraphics[width=1\columnwidth]{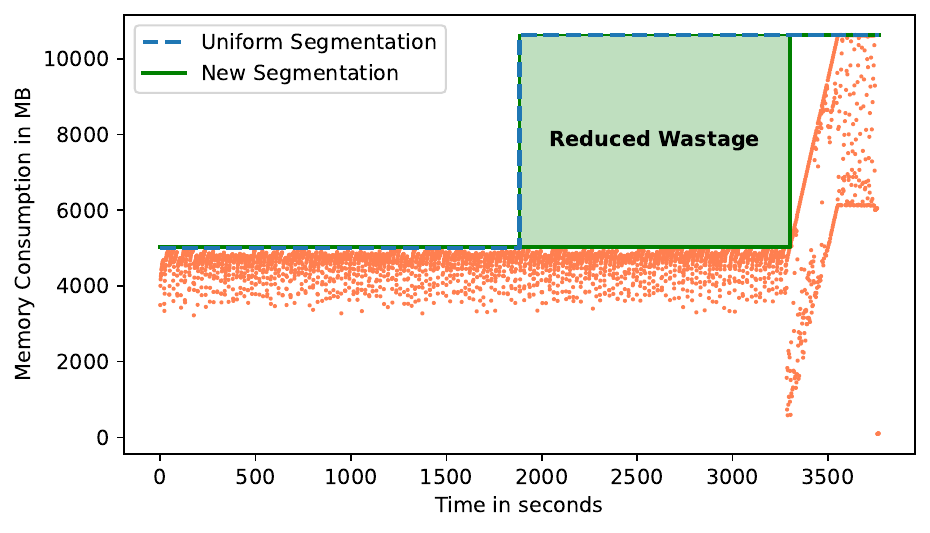}
\caption{Memory consumption over time for a single execution of the BWA task. The dashed blue line represents the prediction with two uniform segments. The green line uses two segments with variable size, generated with our KS+ method.}
\label{fig:segmentation}
\end{figure}

To predict the memory consumption of tasks with variable-size segments over time, we first need a segmentation strategy for a single historical task execution.
The segment parameters, segment start, segment end, and peak memory value per segment, of multiple executions can then be used to build a model that can predict memory usage over time.
Figure~\ref{fig:segmentation} shows the advantage of using two dynamically sized segments (green line) compared to fixed-sized segments (blue line). 
Variable-size segments are able to better model a task's memory consumption and allow for sophisticated retry strategies that involve re-setting the segments for re-execution. 

\subsection{Dynamic Segment Placement} 

\begin{algorithm}[t!]
\caption{Segmentation Algorithm}\label{alg:cap}
\label{alg:segmentation}
\begin{algorithmic}
    \Require \\ $M \gets$ array of memory values \\ $k \gets$ number of output segments
\Function{GetSegments}{$M[\;], k$}
    \State $S \gets [1]$
    \State $P \gets [M_0]$
    \For {$i \gets 1$ to $Length(M)$}
        \If{$M_i \geq P_{-1}$}
            \State $S_{-1} \gets S_{-1} + 1$
        \Else
            \State $Append(P, M_i)$
            \State $Append(S, 1)$
        \EndIf
    \EndFor 

    \While{$Length(P) > k$}
        \State $e \gets [ (P_{i+1} - P_i) \cdot S_i ]$
        \State $idx \gets MinElement(e)$
        \State $S_{idx+1} \gets S_{idx} + S_{idx+1}$
        \State $Remove(S, idx)$
        \State $Remove(P, idx)$
    \EndWhile

    \State \Return {$(S, P)$}
\EndFunction
\end{algorithmic}
\end{algorithm}

We model the memory consumption of a single historical task execution through a step function with $k$ segments, where the i-th segment $1 \leq i \leq k$ is defined by the peak memory consumption $P_i$ in the respective runtime interval of duration $S_i$.
We model the memory consumption as monotonically increasing, i.e. for each segment it must hold that $P_i \leq P_{i+1}$, to avoid task failures caused by reducing memory too early.
The resulting function should define the segments in such a way that the wastage is minimized, i.e., the individual measurement points must be as close as possible to the modeled peak of the corresponding segment, but not higher, as this would result in a task failure.

We developed a greedy segmentation algorithm, which is shown in Algorithm \ref{alg:segmentation}.
The algorithm proceeds in two steps.
First, every data point is considered to be a segment by itself.
The peak memory consumption corresponds to the value of the data point and the size of the segment is one.
We first merge every segment with its predecessor, if the peak value of the segment is smaller than the peak value of the preceding segment.
This is done front to back until the constraint of being monotonically increasing is fulfilled.
In the second step, we further reduce the number of segments.
To decide which segments to merge, we look at the increase of the error $e_i$, which is introduced by merging the $i$-th segment with its successor:
\begin{equation}
e_i = (P_{i+1} - P_{i}) \cdot S_{i} 
\end{equation}

We always merge the segment $S_i$ with the smallest merging error $e_i$, until only $k$ segments are left.
This results in a fast heuristic to find the segments and the corresponding peak values. An example of the resulting segmentation is shown in Figure~\ref{fig:segmentation}.

\subsection{Segment Prediction}

\begin{figure}[t!]
\center
\includegraphics[width=1\columnwidth]{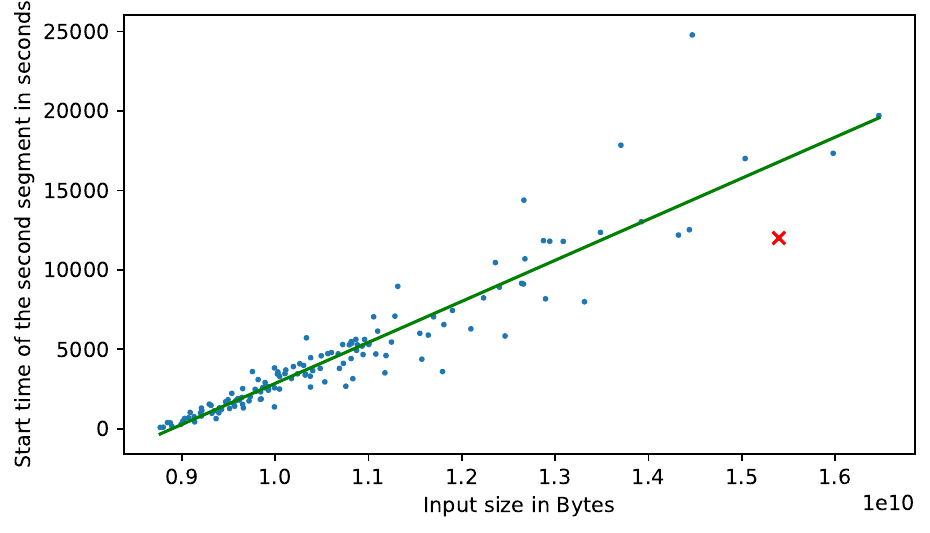}
\caption{Start time for the second segment for all BWA task executions. The blue points represent individual task executions, the green line is the estimation, minimizing the mean squared error. The deviation gets larger for bigger input sizes. The red cross marks an execution that ran much faster than expected.}
\label{fig:segment_prediction}
\end{figure}%

In order to be able to predict the segments and their memory consumption for task execution, Algorithm~\ref{alg:segmentation} is used to gather the $k$ segment parameter pairs $(S_i, P_i)$ for multiple executions of the same task.
As shown in~\cite{da2013toward, da2015online,witt2019feedback, witt2019learning,bader2024lotaru}, the accumulated input file size of all input files is a good indicator for the overall execution time and the peak memory consumption of a task.

We use linear regression based on the input file size $I_t$ for the task execution $t$ to estimate the peak memory consumption $\hat{P}_{t,i}(I_t)$ as well as the start positions for each segment $\hat{S}_{t,i}(I_t)$, resulting in two individual linear regression models for each segment.
These models can then be used to estimate a memory allocation strategy for new executions of the same task with a different input size.
The segment times can scale differently with the input sizes, therefore it is advantageous to use distinct predictions for each segment.
For instance, the execution time of the first process of a task might scale linearly with the input size, while the second process might always take a constant amount of time, resulting in very different time ratios between these two processes for different input sizes.

It is difficult to accurately predict the starting times of the different segments, especially for long-running tasks.
This is due to uncertainty in resource allocation.
For example, a lack of CPU resources caused by other tasks running on the same machine may change the task behavior over time, resulting in slower task execution.
As shown in Figure \ref{fig:segment_prediction}, this makes long-running tasks more susceptible to a larger misprediction.
Since underpredicting the memory results in the restart of a task, it is generally less expensive to overpredict than to underpredict the memory.
Therefore, we added a safety margin by overpredicting the memory peaks by 10\% and underpredicting the segment start times by 15\%.
Since our memory allocation strategy is monotonically increasing, underpredicting the segment start times is always the safer option.

\subsection{Retry Strategy}

\begin{figure}[t!]
\center
\includegraphics[width=1\columnwidth]{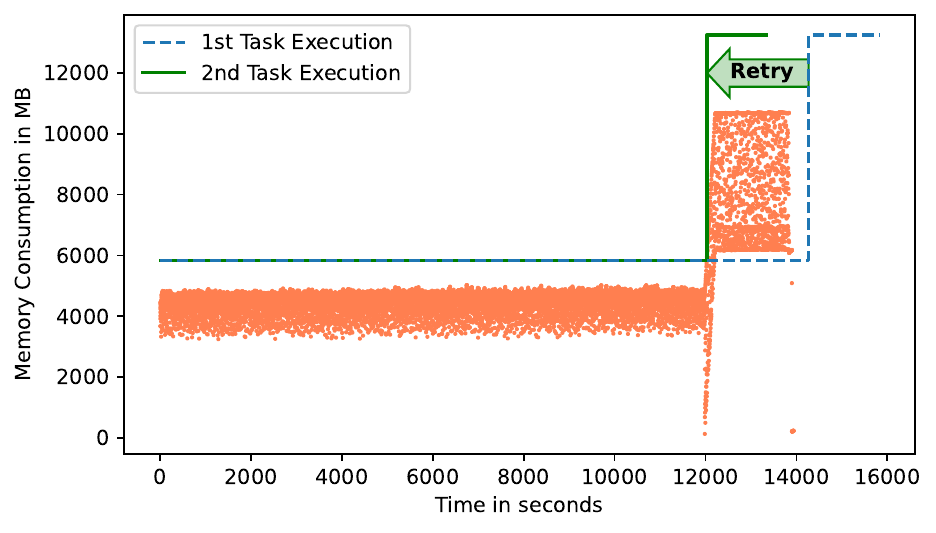}
\caption{Memory consumption over time of the BWA task execution marked in Figure~\ref{fig:segment_prediction}. The dashed blue line represents the predicted KS+ allocation strategy, that fails due to the execution reaching the second segment faster than expected. The green line shows the allocation for the second try. }
\label{fig:retry_strategy}
\end{figure}%

The introduction of dynamic segment sizes enables more flexible retry strategies in case a task execution runs out of memory.
Due to the uncertainty in predicting the segment sizes, it is much more likely that the peak memory consumption was predicted correctly, but the execution reached a more demanding segment earlier than expected.
Instead of doubling the memory allocation upon failure, like most state-of-the-art approaches, we focus on changing the allocation timing.

As soon as a task execution runs out of memory, which is detected, for instance, by the Linux OOM Killer, we check the current runtime of this execution.
This point in time is compared to the expected start time of the next segment.
The start times of all succeeding segments are reduced by a factor so that the next segment would have started when the execution ran out of memory.
This approach is visualized in Figure~\ref{fig:retry_strategy}.
Only if the point of failure is already in the last segment, the peak of the segment is increased by 20\%.
If the task fails again, this procedure is repeated until the execution finishes successfully.
This strategy utilizes the knowledge about when the execution ran out of memory to drastically reduce the uncertainty in the segment start time prediction for the current execution.

\section{Evaluation}\label{sec:Eval1}
Our evaluation section contains the experimental setup, briefly explains the baselines, and shows the experimental results.

\subsection{Experimental Setup}

For our experiments, we use the data published alongside the original k-Segments method~\cite{baderDiedrichPredicting2023}.
The data consists of two nf-core~\cite{ewels2020nf} workflows, eager~\cite{yates2021reproducible} and sarek~\cite{hanssen2024scalable}.
Eager uses input data from a 2018 study that examines the population history in the area of the Eurasian steppe~\cite{damgaard2018137}.
The sarek workflow was run with input from a 2022 study, which investigated estrogen receptor mutations in breast cancer~\cite {harrod2022genome}.
Both workflows were executed on a machine equipped with an AMD EPYC 7282 processor and 128GB DDR4 memory.

\begin{figure}[t!]
	\centering
	\includegraphics[width=0.9\columnwidth]{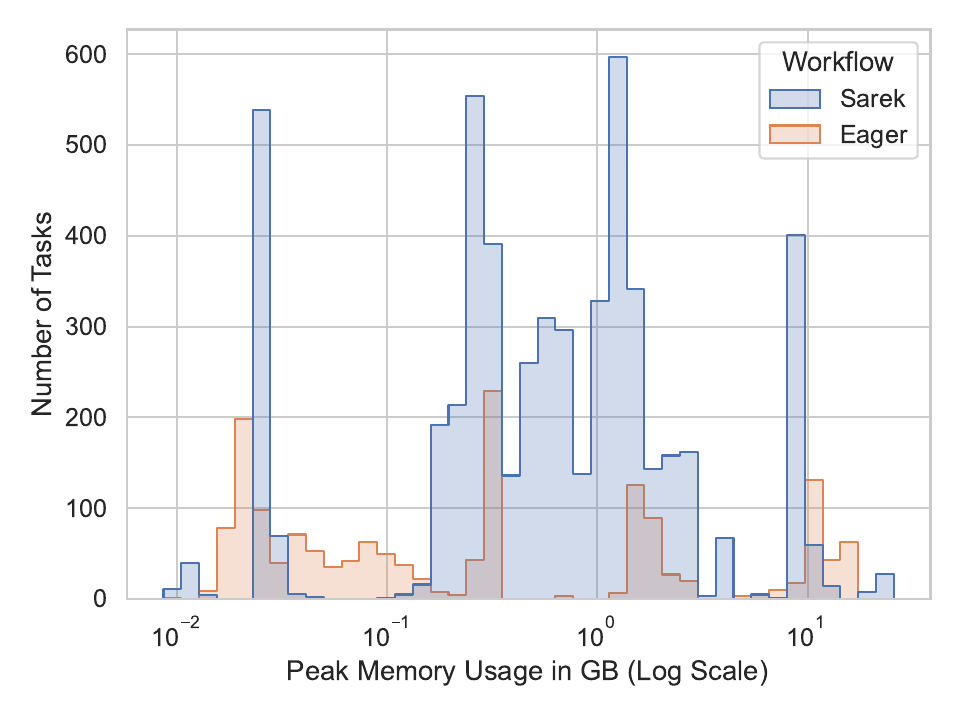}
	\caption{Memory consumption of the experimental workflow tasks.}
	\label{fig:experimental_data}
\end{figure}

Figure~\ref{fig:experimental_data} provides an overview of the two workflows and their peak memory consumption.
It can be observed that the sarek workflow has more task instances and an average peak memory usage of 1.67 GB.
In contrast, the eager workflow contains fewer task instances but has an average peak memory usage of 2.31 GB.


We run the experiments ten times, each time with a different seed that is used to split the train/test data.
The values given are averages over all runs.
For wastage, the metric used is gigabyte seconds (GBs), which accounts for memory wastage over time. 
The wastage for a single task execution can be defined as the difference between requested and used memory over time plus the sum of allocated memory over time from its failed task executions and thus also accounts for task failures.

\subsection{Baselines}

We compare our KS+ method with four state-of-the-art methods from the literature and the workflow developers' default as a sanity baseline.

Tovar et al.~\cite{tovar2017job} predict a task's peak memory using historical peak memory probabilities (\textit{Tovar-PPM}).
If a task fails due to an initial under-allocation, the maximum available memory of the machine is allocated for re-execution.
We use their publicly available source code as the basis for implementation in our simulation environment.

We provide an improved version of Tovar-PPM, \textit{PPM-Improved}, which does not allocate the maximum available memory upon task failure, but instead doubles it, resulting in potentially less wastage for cluster setups with nodes equipped with lots of memory.

\textit{k-Segments Selective} and \textit{k-Segments Partial}~\cite{baderDiedrichPredicting2023} also serve as baselines and dynamically predict memory wastage over time using equally-sized segments and a failure strategy that either offsets everything after the failed segment or selectively offsets only the failed segment. 

The default baseline uses the workflow developers' task memory limits.

\subsection{Experimental Results}

\begin{figure*}[t!]
\centering
\begin{subfigure}[t]{\columnwidth}
   \centering
  \includegraphics[width=1\columnwidth]{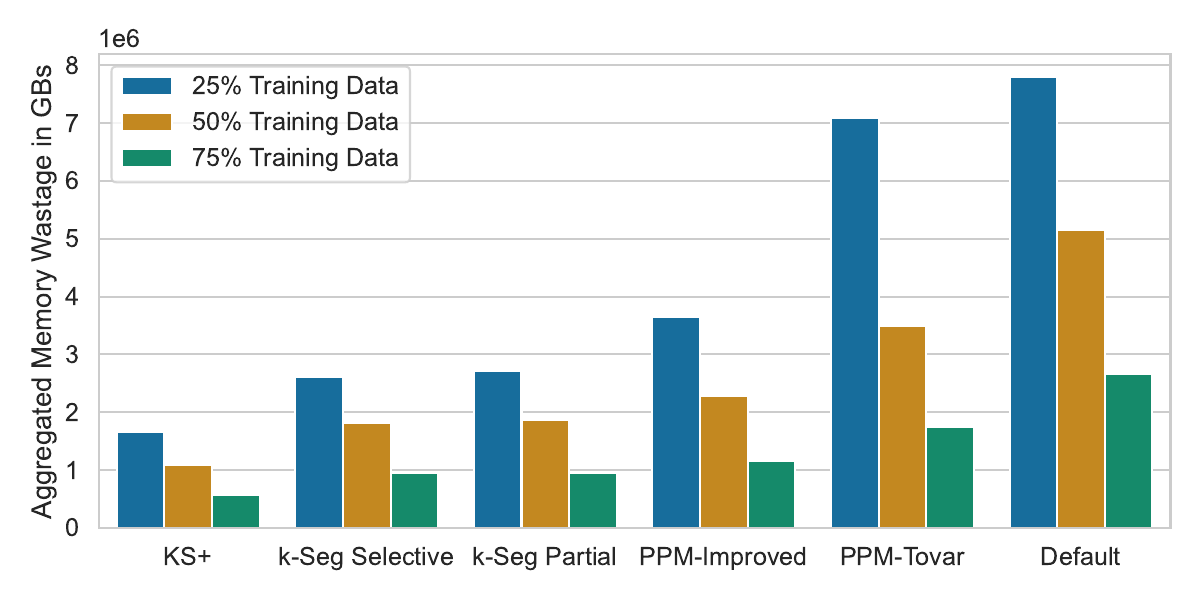}
  \caption{Eager workflow}
  \label{fig:eval_aggregated_eager}
\end{subfigure}
\begin{subfigure}[t]{\columnwidth}
  \centering
  \includegraphics[width=1\columnwidth]{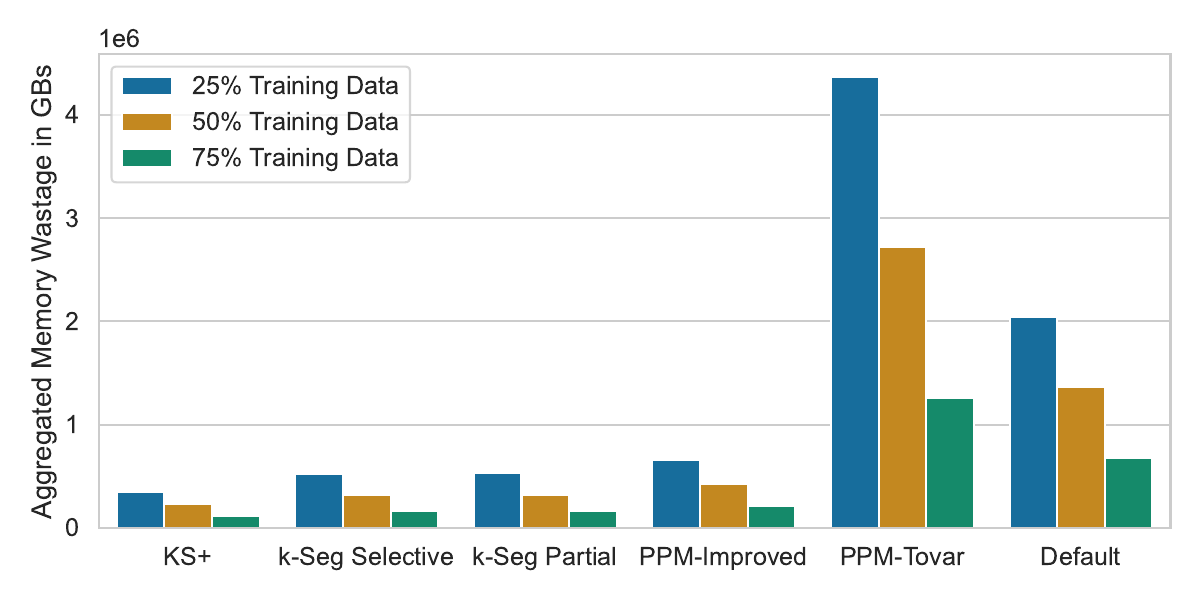}
  \caption{Sarek workflow}
  \label{fig:eval_aggregated_sarek}
\end{subfigure}
\caption{Aggregated memory wastage in GBs for each method.}
\label{fig:eval_aggregated}
\end{figure*}%

\begin{figure*}[t!]
\centering
\begin{subfigure}[t]{\columnwidth}
   \centering
  \includegraphics[width=1\columnwidth]{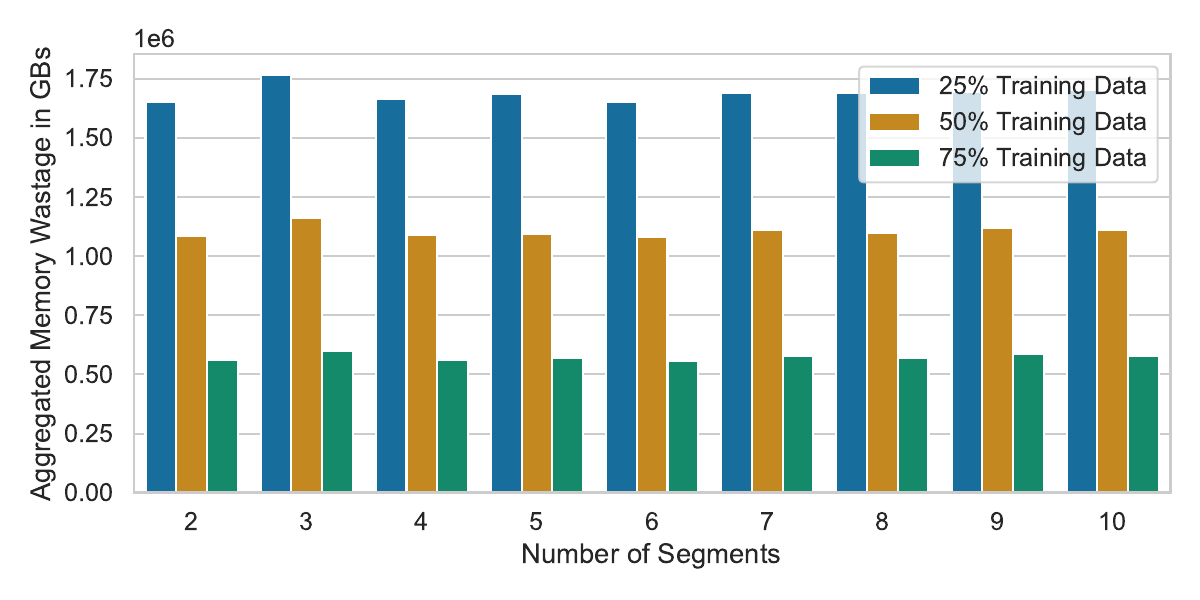}
  \caption{Eager workflow}
  \label{fig:eval_segments_eager}
\end{subfigure}
\begin{subfigure}[t]{\columnwidth}
  \centering
  \includegraphics[width=1\columnwidth]{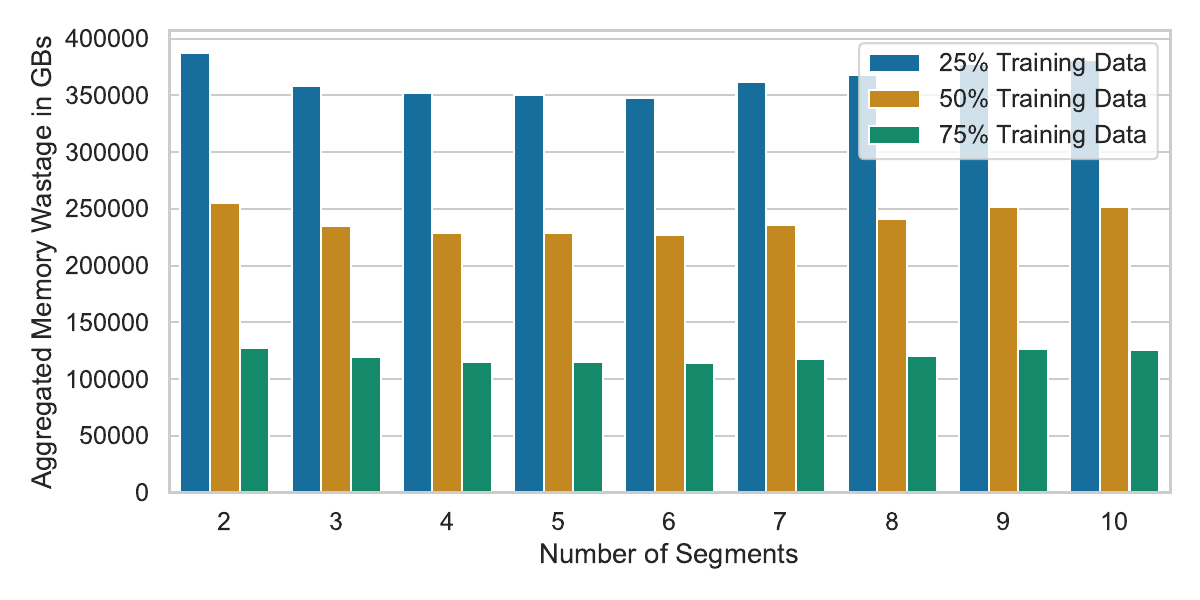}
  \caption{Sarek workflow}
  \label{fig:eval_segments_sarek}
\end{subfigure}
\caption{Effect of the number of segments on the memory wastage of our KS+ method.}
\label{fig:eval_segments}
\end{figure*}%

In this section, we first compare the aggregated memory wastage for all methods, followed by a task-based analysis.
Subsequently, we analyze the impact of the number of segments on the prediction results of our KS+ method.

Figure~\ref{fig:eval_aggregated} shows the aggregated memory wastage for each method.
For the eager workflow, Figure~\ref{fig:eval_aggregated_eager}, KS+ achieves the lowest memory wastage among all methods.
We can observe a reduction in memory wastage of 36\%, 39\%, and 40\% for 25\%, 50\%, and 75\% training data compared to the best-performing baseline and a reduction of  54\%, 52\%, and 51\% compared to the best-performing peak memory prediction baseline, PPM-Improved.
Also, for the sarek workflow, Figure~\ref{fig:eval_aggregated_sarek}, KS+ significantly outperforms all baselines.
Compared to the best-performing baseline, k-Segments Selective, it can be observed that KS+ reduces the wastage by 31\%, 28\%, and 29\%.
Compared to PPM-Improved, there is an average reduction of 45\% for all training sets. 
Notably, PPM-Improved significantly outperforms PPM-Tovar, with the retry strategy adjustment being the only changed feature.
We suspect that this is due to the large amount of memory on our experimental nodes, which results in high memory wastage for retried task executions.
This could also cause the default baseline to achieve lower memory wastage than PPM-Tovar for the Sarek workflow.

\begin{figure*}[t!]
	\centering
	\includegraphics[width=1\textwidth]{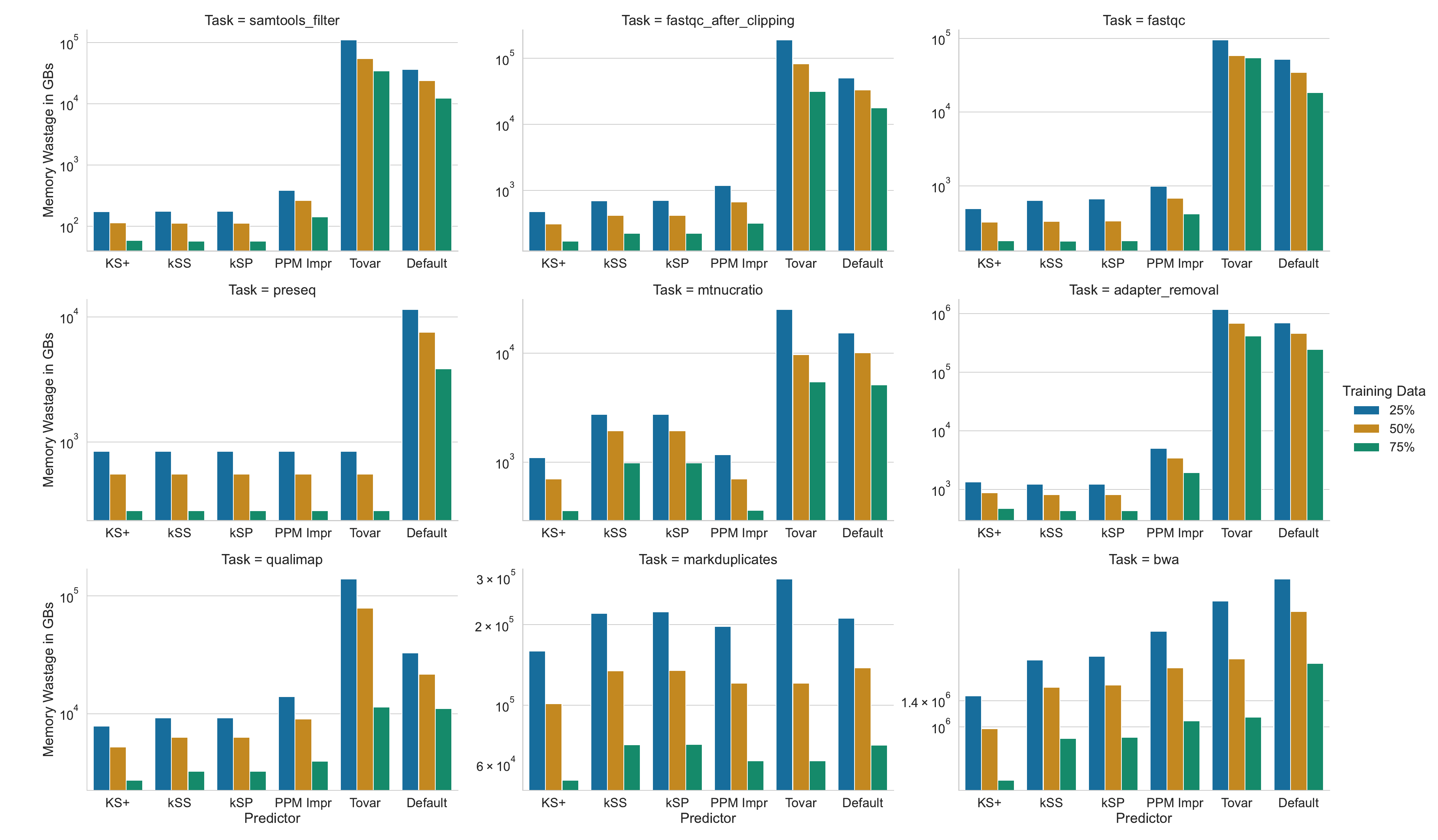}
	\caption{Memory wastage in GBs for each of the nine tasks to be predicted in the eager workflow using different amounts of training data.}
	\label{fig:eval_eager_tasks}
\end{figure*}

Figure~\ref{fig:eval_eager_tasks} shows the memory wastage for each task in the eager workflow.
It can be observed that the bwa task contributes the most to the total memory wastage.
For this task, we observe a memory wastage reduction of 42\%, 41\%, and 37\% compared to the best-performing baseline.
For the mtnucratio task, we observe the largest relative reduction in memory wastage among all eager tasks.
The AdapterRemoval task and the Samtools task are the only two tasks that show a slight increase in memory wastage compared to the k-Segments Selective method. 
Over both workflows, we achieve an unweighted average memory wastage reduction of 20\%, 18\%, and 21\% compared to k-Segments Selective and a reduction of 37\%, 38\%, and 40\% compared to PPM-Improved.

Finally, we analyze how the number of segments affects KS+'s memory wastage.
Figure~\ref{fig:eval_segments} shows the aggregated memory wastage as a function of the number of segments.
For both workflows, there are no significant outliers and KS+ achieves low memory wastage.
Furthermore, we cannot observe that fewer or more segments have a positive or negative impact on the total memory wastage.
Both plots show minimal wastage when six segments are selected.
However, not all tasks follow this trend, and different workflows and tasks may benefit from a different selection.

\section{Related Work}\label{sec:RELATED_WORK}
This section first describes research on workflow task memory prediction and then compares it to our KS+ method.

Tovar et al.~\cite{tovar2017job} introduced a task memory prediction strategy for high-throughput scientific workflows. 
They propose a model that predicts the peak resource usage for a specific task using an analytical approach. 
The model can be adjusted to achieve two different objectives: maximizing throughput or minimizing resource wastage. 
Both objectives optimize resource usage based on a slow-peaks model, which assumes a scenario where tasks fail at the end of their execution. 

Witt et al.~\cite{witt2019feedback} presented an online feedback loop-based resource allocation method to reduce resource wastage. 
Their predictor uses a task's aggregated input file sizes and employs various linear regression models to predict peak memory usage, adjusting the linear regression to overpredict and thus prevent task failures due to underprovisioning. 
They propose several offset strategies: the LR mean~± strategy considers the standard deviation as an offset, the LR mean - strategy considers only negative prediction errors, and the LR max strategy adds the largest observed underprediction as an offset. 

Witt et al.~\cite{witt2019learning} propose a second method to address the memory allocation problem by focusing on minimizing resource wastage rather than prediction error. 
Again, they assume a relationship between the size of the input data and a task's peak memory usage, training a linear model based on this assumption. 
They evaluate different strategies for handling task failures and show that the right failure-handling strategy has a significant impact on resource wastage because underpredictions are hard to avoid.

In our own previous work~\cite{bader2022leveraging}, we introduced two reinforcement learning approaches using gradient bandits and Q-learning to minimize resource wastage. 
These reinforcement learning bandits did not implement an offset technique as the reinforcement learning agents are inherently incentivized to avoid underpredictions. 
One drawback of our proposed methods is that they do not account for the dependency between task input size and resource usage.

We also presented a method to predict the memory consumption of workflow tasks over time~\cite{baderDiedrichPredicting2023}.
We utilized time series monitoring data to predict the expected task runtime, which we then divided into equally sized segments.
For each segment, we trained a linear regression model that predicts peak memory, resulting in a step function that models a task's memory consumption over time.
Both the internal runtime prediction model and the segment-by-segment peak memory prediction models incorporate an offset strategy.

Compared to the static peak memory prediction methods from the literature, our KS+ method dynamically predicts memory over time, allowing for changing memory allocation during workflow execution and potentially reducing memory waste.
Since such fine-grained memory prediction opens the space for underpredictions and subsequent task failures, we extend our previously presented k-Segments model~\cite{baderDiedrichPredicting2023} with a) dynamic segment size selection, b) an improved memory prediction model for the segments, and c) an improved failure handling that takes advantage of our dynamically sized segment lengths.

\section{Conclusion}\label{sec:CONCLUSION}
This paper presented KS+, a method for dynamically predicting workflow task memory consumption over time.
To this end, KS+ predicts segments that define the timing and length of memory allocations.
It then predicts the memory usage for each of these segments.
Our method further employs a dynamic re-assignment strategy that not only adjusts the memory upon task failure but also re-assigns the segments. 

Our experimental evaluation with four state-of-the-art baselines and two workflows from the popular nf-core repository has shown an average memory wastage reduction of 38\% compared to the best state-of-the-art baseline. 
Furthermore, we have shown that our model is robust to different segment configurations.

In the future, we plan to dynamically determine the optimal number of segments for each task.

\section*{Acknowledgments}
{\small Funded by the Deutsche Forschungsgemeinschaft (DFG, German Research Foundation) as FONDA (Project 414984028, SFB 1404).}

\bibliographystyle{IEEEtran}
\balance
\bibliography{references}

\end{document}